\documentclass{article}
\usepackage{spconf,amsmath,graphicx}

\usepackage{bm}
\usepackage{textcomp}
\usepackage{amsfonts}
\usepackage{psfrag}
\usepackage{epsfig,epstopdf,amssymb}
\usepackage{stfloats,multirow,booktabs,color,ifpdf,float,array}
\usepackage{algorithm,algorithmic}
\usepackage{tabu}

\title{Unsupervised Model-based speaker adaptation of \\ end-to-end lattice-free MMI model for speech recognition}
\name{Xurong Xie$^{1}$, Xunying Liu$^{2}$, Hui Chen$^{1}$, Hongan Wang$^{1}$
\thanks{This research was supported by National Key R\&D Program of China (2020YFC2004100) and National Natural Science Foundation of China Grant (NSFC 62106255).}}
\address{
  $^1$Institute of Software, Chinese Academy of Sciences, China\\
  $^2$The Chinese University of Hong Kong, Hong Kong SAR, China\\
  {\tt \normalsize xurong@iscas.ac.cn, xyliu@se.cuhk.edu.hk, chenhui@iscas.ac.cn, hongan@iscas.ac.cn}}
\begin{document}
\ninept
\maketitle
\begin{abstract}
Modeling the speaker variability is a key challenge for automatic speech recognition (ASR) systems. In this paper, the learning hidden unit contributions (LHUC) based adaptation techniques with compact speaker dependent (SD) parameters are used to facilitate both speaker adaptive training (SAT) and unsupervised test-time speaker adaptation for end-to-end (E2E) lattice-free MMI (LF-MMI) models. An unsupervised model-based adaptation framework is proposed to estimate the SD parameters in E2E paradigm using LF-MMI and cross entropy (CE) criterions. Various regularization methods of the standard LHUC adaptation, e.g., the Bayesian LHUC (BLHUC) adaptation, are systematically investigated to mitigate the risk of overfitting, on E2E LF-MMI CNN-TDNN and CNN-TDNN-BLSTM models. Lattice-based confidence score estimation is used for adaptation data selection to reduce the supervision label uncertainty. Experiments on the 300-hour Switchboard task suggest that applying BLHUC in the proposed unsupervised E2E adaptation framework to byte pair encoding (BPE) based E2E LF-MMI systems consistently outperformed the baseline systems by relative word error rate (WER) reductions up to 10.5\% and 14.7\% on the NIST Hub5'00 and RT03 evaluation sets, and achieved the best performance in WERs of 9.0\% and 9.7\%, respectively. These results are comparable to the results of state-of-the-art adapted LF-MMI hybrid systems and adapted Conformer-based E2E systems.
\end{abstract}

\setlength{\parskip}{1pt plus 1pt minus 1pt}
\setlength{\intextsep}{1pt plus 1pt minus 1pt}
\setlength{\textfloatsep}{4pt plus 1pt minus 1pt}
\setlength{\dblfloatsep}{3pt plus 1pt minus 1pt}
\setlength{\dbltextfloatsep}{3pt plus 1pt minus 1pt}
\setlength{\abovecaptionskip}{4pt plus 1pt minus 1pt}
\setlength{\abovedisplayskip}{1pt plus 1pt minus 1pt}
\setlength{\belowdisplayskip}{1pt plus 1pt minus 1pt}

\section{Introduction}
\label{sec:intro}

Recently, there is a major trend in the ASR field that more attentions are received by the end-to-end (E2E) modeling approaches. They typically aim to train a single neural network acoustic model in one stage without using alignments, state-tying decision trees, and well-trained GMM-HMMs. These methods are represented by connectionist temporal classification (CTC) \cite{graves2006connectionist}, attention-based encoder-decoder (AED) \cite{chorowski2015attention} (or listen, attend and spell (LAS) \cite{chan2016listen}), recurrent neural network transducer (RNN-T) \cite{graves2012sequence}, Transformer \cite{dong2018speech}, and convolution-augmented Transformer (Conformer) \cite{gulati2020conformer}. Alternatively, lattice-free maximum mutual information (LF-MMI), which is widely used in the hybrid/DNN-HMM paradigms \cite{Povey2016Purely}, can be employed for E2E modeling \cite{hadian2018end,hadian2018flat} in a flat-start manner using phoneme-based or character-based setting. The E2E LF-MMI models have shown comparable performance to the LF-MMI time-delay neural network (TDNN) hybrid models, and to other popular E2E models while often using significantly smaller models.

A key challenge for hybrid and E2E ASR systems is to reduce the mismatch against target users resulted from the systematic and latent variation in speech. Speaker level characteristics are a major source of such variability, which represents factors such as accent, age, gender, and other idiosyncrasies or physiological differences. Speaker adaptation techniques for current DNN acoustic modeling can be broadly characterized into three categories. In auxiliary speaker embedding-based techniques, a compact vector, for examples, i-vector~\cite{saon2013speaker} or speaker code~\cite{abdel2013fast}, is used to encode speaker-dependent (SD) characteristics and augment standard acoustic front-ends.
In feature transformation-based techniques, affine transforms \cite{leggetter1995maximum,digalakis1995speaker,gales1998maximum} (e.g., feature-space maximum likelihood linear regression (f-MLLR)) or vocal tract length normalization \cite{lee1996speaker,uebel1999investigation} applied to the acoustic front ends are used to produce speaker-independent (SI) features.
In model-based techniques, SD parameters represented by learning hidden unit contributions (LHUC) \cite{swietojanski2016learning,swietojanski2015differentiable}, parameterized hidden activation functions \cite{zhang2016dnn}, various linear transforms \cite{neto1995speaker,gemello2007linear,li2010comparison}, and interpolation techniques \cite{wu2015multi,tan2016cluster} are separately or jointly estimated with the remaining DNN parameters in test-time adaptation or speaker adaptive training (SAT) \cite{anastasakos1996compact,swietojanski2016learning}, respectively.
These works were predominantly conducted for hybrid systems.

Relatively limited number of works on speaker adaptation were conducted for E2E systems.
For examples, speaker embeddings (e.g., i-vectors) can be exploited directly as auxiliary input features \cite{tuske2021limit} or to build an attention-based SD component \cite{fan2019speaker} in AED and Transformer models. The f-MLLR is also popular for E2E systems \cite{chorowski2014end,tomashenko2018evaluation}.
Moreover, model-based techniques were applied to CTC \cite{li2018speaker}, AED \cite{ochiai2018speaker,weninger2019listen}, Transformer \cite{deng2022confidence}, Conformer \cite{deng2022confidence,huang2021rapid}, or RNN-T \cite{huang2020rapid} based E2E models. The SD parameters can be the whole network, parts of the network, or additional SD components such as linear hidden networks (LHN) or LHUC applied to encoder, attention or decoder modules.
However, none of these speaker adaptation works is conducted for E2E LF-MMI models.

For E2E models, there are two major challenges that efforts on developing model-based speaker adaptation techniques are confronted with.
First, the often limited amounts of speaker specific data requires highly compact and regularized SD parameter representations to be used to mitigate the risk of overfitting during adaptation, for example, LHUC-based adaptation techniques making use of relatively lower dimension of SD parameters compared to the techniques using full linear transforms or the whole networks. Moreover, the Bayesian adaptation \cite{xie2019blhuc,xie2021bayesian,deng2022confidence}, Kullback-Leibler (KL) divergence regularization~\cite{yu2013kl,li2018speaker,weninger2019listen} and maximum a posteriori (MAP) adaptation~\cite{huang2015maximum} also have the potential to reduce overfitting by regularizing parameter uncertainty.
Second, SD parameter estimation may be sensitive to the underlying supervision uncertainty caused by the error introduced in unsupervised test-time manner of model-based adaptation. For instance, the LHUC adaptation requires an initial decoding pass over the adaptation data to produce initial speech transcription as supervision for the subsequent SD parameter estimation. This issue is particularly prominent with E2E systems that directly learn the surface word or token sequence labels in the adaptation supervision, as opposed to the frame-based phonetic state alignments in hybrid systems. In \cite{deng2022confidence}, a confidence estimation module is trained to selected the data with higher accuracy for adaptation.

In order to address these issues of model-based speaker adaptation for E2E LF-MMI models, in this paper, the LHUC-based adaptation techniques with compact SD parameters are used to facilitate both SAT and unsupervised test-time speaker adaptation. We propose an unsupervised adaptation framework to estimate the SD parameters in E2E paradigm using LF-MMI and cross entropy (CE) criterions. In addition to standard LHUC adaptation, various regularization methods including the Bayesian LHUC (BLHUC), MAP-LHUC, and KL-LHUC are investigated in this E2E adaptation framework to mitigate the risk of overfitting, on E2E LF-MMI CNN-TDNN and CNN-TDNN-BLSTM models. In order to reduce the supervision uncertainty during adaptation, lattice-based recognition hypotheses are utilized to estimate the confidence score \cite{evermann2000posterior,mangu2000finding} for selecting the most ``trustworthy'' subset of speaker specific data.

Experiments on the 300-hour Switchboard corpus suggest that applying BLHUC in the proposed E2E adaptation framework to E2E LF-MMI systems consistently outperformed the baseline systems by relative word error rate (WER) reductions up to 11.1\% and 14.7\% on the NIST Hub5'00 and RT03 evaluation sets, respectively.
By joint use of E2E BLHUC adaptation and SAT, the best byte pair encoding (BPE) based E2E LF-MMI CNN-TDNN-BLSTM system using i-vector and cross-utterance Transformer and LSTM language models (LMs) rescoring obtained WERs of 9.0\% and 9.7\% on the Hub5'00 and RT03 evaluation sets, respectively. These results are comparable to both the results of state-of-the-art adapted LF-MMI hybrid systems and the state-of-the-art adapted Conformer-based E2E systems.
To the best of our knowledge, this is the first systematic investigation work on model-based speaker adaptation for E2E LF-MMI models using both phoneme-based and BPE-based settings.

The rest of this paper is organized as follows. Section \ref{sec:lfmmi} briefly reviews the E2E LF-MMI model. LHUC-based techniques using the proposed E2E adaptation framework are introduced in Section \ref{sec:adapt}. Section \ref{sec:exps} presents the experiments and results on the 300-hour Switchboard task. Finally, the conclusions are drawn in Section \ref{sec:concl}.

\section{End-to-end LF-MMI}
\label{sec:lfmmi}

\begin{figure}[!htbp]
  \begin{center}
    \includegraphics[width=8.5cm,height=1.3cm]{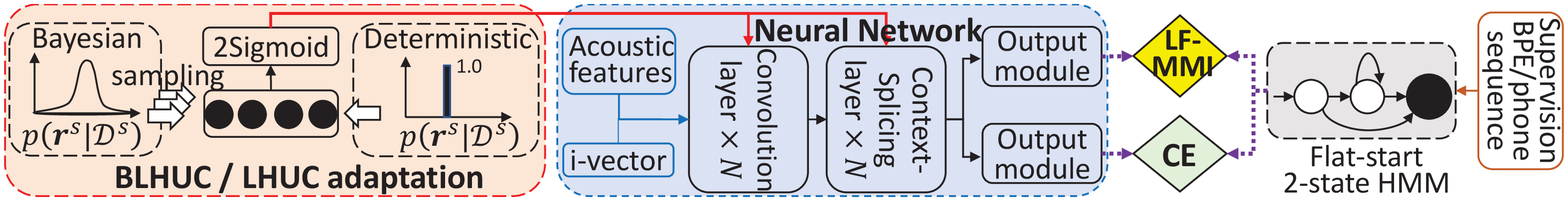}
    \vspace{-1.0em}
    \caption{The E2E adaptation framework for E2E LF-MMI model. }
    \vspace{-1.0em}
    \label{fig:e2e}
  \end{center}
\end{figure}

The end-to-end (E2E) lattice-free maximum mutual information (LF-MMI) model \cite{hadian2018end,hadian2018flat} is actually a neural network trained with LF-MMI criterion using flat-start HMMs with uniform and fixed transition probabilities. The purpose of using HMM is to constraint the possible state sequences by its (commonly two-state) topology for E2E model training and decoding. An HMM is built to represent each bi-phone without state tying for phoneme-based setting, and represent each BPE token for BPE-based setting.

The MMI is a discriminative objective function aiming to maximize the probability of reference word sequences while minimizing the probability of all other word sequences. Given the data set $\mathcal{D}=\{{\bm O},{\bm W}\}$ with acoustic features ${\bm O}$ and the corresponding word sequences ${\bm W}$, LF-MMI is to maximize the objective function
\begin{eqnarray}
\mathcal{F}_{\!L\!F\!-\!M\!M\!I}\!(\mathcal{D}) \!\! = \!\! \sum_{u} \! \log \!\! \frac{P(\!{\bm o}_u|{\bm w}_u\!)\!P(\!{\bm w}_u\!)}{\sum_{\tilde{{\bm w}}} \! P(\!{\bm o}_u|\tilde{{\bm w}}\!)\!P(\!\tilde{{\bm w}}\!)} \!\! \approx \!\! \sum_{u} \! \log \!\! \frac{P(\!{\bm o}_u|\mathbb{M}_{num}^{(u)}\!)}{P(\!{\bm o}_u|\mathbb{M}_{den}\!)} \!\!\!
\label{eq:mmi}
\end{eqnarray}
where ${\bm o}_u$ denotes the acoustic feature sequence of the $u$th utterance in the data set, and $\mathbb{M}_{num}^{(u)}$ and $\mathbb{M}_{den}$ are the numerator graph and denominator graph containing all possible HMM state sequences given the word sequence ${\bm w}_u$ in the numerator and given all word sequences in the denominator, respectively. Both the $\mathbb{M}_{num}^{(u)}$ and $\mathbb{M}_{den}$ are computed as finite-state transducers for E2E training.
To make the computation of $\mathbb{M}_{den}$ more efficient and tractable, an N-gram token-based language model (LM) is estimated using the reference word sequences and forward-backward algorithm is exploited.
A 4-gram mono-phone LM is estimated for phoneme-based setting, and a 3-gram BPE LM is estimated for BPE-based setting.

Further, an additional minimum cross entropy (CE) criterion is often employed to smooth the training. This is given by
\begin{eqnarray}
\mathcal{F}_{C\!E}(\mathcal{D}) = - \sum\nolimits_{u,t,y_t} p(y^u_t|\mathbb{M}_{num}^{(u)}) \log P_{C\!E}(y^u_t|{\bm o}_{u}) \label{eq:ce}
\end{eqnarray}
where $p(y^u_t|\mathbb{M}_{num}^{(u)})$ is the occupation probability of state $y^u_t$ at time $t$ for utterance $u$ given the numerator graph $\mathbb{M}_{num}^{(u)}$.
In practice, $P_{C\!E}(y^u_t|{\bm o}_{u})$ is commonly implemented as different network output module from that for LF-MMI criterion, as the blue block in Fig \ref{fig:e2e}.

\section{End-to-end Adaptation Framework}
\label{sec:adapt}

The unsupervised test-time adaptation of model-based adaptation techniques originally proposed for hybrid model requires an initial decoding pass over the adaptation data to produce frame-based alignment of state sequences as supervision labels for SD parameter estimation. Although state alignments can also be produced by decoding pass of E2E LF-MMI model, using state alignments for E2E model adaptation may introduce some mismatch as the model is not trained with state alignments. To address the issue, we propose an E2E framework of model-based adaptation by modifying, for example, the LHUC-based adaptation techniques to use E2E LF-MMI and CE for SD parameter estimation without using state alignment. In this E2E framework, the word sequences produced by initial decoding over the adaptation data are utilized for SD parameter estimation.

\subsection{LHUC adaptation}
\label{sec:lhuc}

The key idea of LHUC~\cite{swietojanski2016learning,swietojanski2015differentiable} speaker adaptation is learning a set of SD scaling vectors to modify the amplitudes of neural network hidden unit activations for each speaker, as the red block shown in Fig \ref{fig:e2e}. In a hidden layer of a network, let ${\bm r}^{s}$ denotes the parameters of a scaling vector for speaker $s$, without loss of generality, the hidden unit activations using LHUC adaptation is expressed as
${\bm h}^{s} = \xi({\bm r}^{s}) \otimes {\bm h}$,
where $\otimes$ denotes the Hadamard product, and ${\bm h}$ denotes the unadapted hidden unit activations. In this paper, ${\bm h}$ is activated by ReLU function and the scaling vector is modeled by element-wise activation function $\xi(\cdot)=2\text{Sigmoid}(\cdot)$ with range $(0,2)$.

The standard LHUC adaptation for hybrid model estimates the SD parameters ${\bm r}^{s}$ by minimzing alignment-based CE criterion. For E2E LF-MMI model, we use the loss function interpolating the LF-MMI and CE criterions for adaptation. Given the adaptation data $\mathcal{D}^s = \{{\bm O}^s,{\bm W}^s\}$ for speaker $s$, ${\bm r}^{s}$ is optimized by minimizing
\begin{eqnarray}
\mathcal{L}_{L\!H\!U\!C}(\mathcal{D}^s|{\bm r}^{s}) \label{eq:mmice_lhuc}  = - \gamma_1 \mathcal{F}_{L\!F\!-\!M\!M\!I}(\mathcal{D}^s|{\bm r}^{s}) \!+\! \gamma_2 \mathcal{F}_{C\!E}(\mathcal{D}^s|{\bm r}^{s})
\label{eq:mmice_lhuc}
\end{eqnarray}
where $\gamma_1$ and $\gamma_2$ are interpolation scales for LF-MMI and CE criterions given by equations (\ref{eq:mmi}) and (\ref{eq:ce}) respectively. Given the numerator graph $\mathbb{M}_{num}^{(s,u)}$ generated by word sequence ${\bm w}^s_u$, the gradient over $r^{s}_d$ on the $d$th dimension for optimization is computed as
\begin{eqnarray}
\!\!\!\!\!\!\!&&\!\!\!\!\!\!\! \frac{\partial \mathcal{L}_{L\!H\!U\!C}(\mathcal{D}^s|{\bm r}^{s})}{\partial r^{s}_d} = \label{eq:grad_lhuc} \\
\!\!\!\!\!\!\!&&\!\!\!\!\!\!\! \sum\nolimits_{u,t,y^u_t} \!\!\! \left( \! \gamma_1(P(y^u_t|{\bm o}^s_{u},\mathbb{M}_{den},{\bm r}^{s}) \!-\! P(y^u_t|\mathbb{M}_{num}^{(s,u)})) \frac{\partial a^{L\!F\!-\!M\!M\!I}_{s,u,t,y^u_t}}{\partial h^{s}_{u,t,d}} + \right. \nonumber \\
\!\!\!\!\!\!\!&&\!\!\!\!\!\!\! \left. \gamma_2(P_{C\!E}(y^u_t|{\bm o}^s_{u},{\bm r}^{s}) - P(y^u_t|\mathbb{M}_{num}^{(s,u)})) \frac{\partial a^{C\!E}_{s,u,t,y^u_t}}{\partial h^{s}_{u,t,d}} \right) h_{u,t,d} \frac{\partial \xi(r^{s}_d)}{\partial r^{s}_d} \nonumber
\end{eqnarray}
where $P(y^u_t|{\bm o}^s_{u},\mathbb{M}_{den},{\bm r}^{s})$ is the posterior probability of state $y^u_t$ at time $t$ for utterance $u$ given acoustic features ${\bm o}^s_{u}$ computed over the denominator graph $\mathbb{M}_{den}$, and $a^F_{s,u,t,y^u_t}$ denotes the $y^u_t$th value of output module for $F\in\{L\!F\!-\!M\!M\!I,C\!E\}$ preceding Softmax function.
For updating the SD parameters in E2E framework, we modify each utterance length of adaptation data to the nearest one of around 40 distinct lengths by simply extending the wave data with silence, as the utterances cannot be split into chunks without alignment.

\subsection{Bayesian LHUC adaptation}
\label{sec:blhuc}

The limited data amount for adaptation may lead to uncertainty of SD parameters in LHUC adaptation performing deterministic parameter estimation. To address the issue, Bayesian LHUC (BLHUC) adaptation \cite{xie2019blhuc,xie2021bayesian} explicitly models the uncertainty by using posterior distribution of ${\bm r}^{s}$ estimated with variational approximation $q_s({\bm r}^{s})$. Prediction for test data is given by
$P(\hat{{\bm Y}}^s|\hat{{\bm O}}^s,\mathcal{D}^s) = \int P(\hat{{\bm Y}}^s|\hat{{\bm O}}^s,{\bm r}^{s})q_s({\bm r}^{s}) d{\bm r}^{s} \approx P(\hat{{\bm Y}}^s|\hat{{\bm O}}^s,\mathbb{E}_{q_s}[{\bm r}^{s}])$.

In \cite{xie2019blhuc,xie2021bayesian} the variational bound is derived from marginal CE using state alignments. For CE criterion in the E2E adaptation framework, variational bound can be derived from similar marginalization
\begin{eqnarray}
\!\!\!\! && \!\!\!\!\!\!\!\! - \!\!\! \sum\nolimits_{{\bm Y}} p_{n^s}^{({\bm Y})} \! \log \! P_{C\!E}({\bm Y}|{\bm O}^s) \! = \! - \!\!\! \sum\nolimits_{{\bm Y}} p_{n^s}^{({\bm Y})} \! \log \!\!\! \int \!\!\! P_{C\!E}({\bm Y}\!,\! {\bm r}^{s}|{\bm O}^s) d{\bm r}^{s} \nonumber \\
\!\!\!\! && \!\!\!\!\!\!\!\! \leq  - \!\!\! \sum\nolimits_{u,t,y^u_t}  p_{n^{s}_u}^{(y^u_t)} \!\!\! \int \!\!\! q_s({\bm r}^{s}) \! \log \! P_{C\!E}(y^u_t|{\bm o}^s_{u},{\bm r}^{s}) d{\bm r}^{s} \!+\! K\!L(q_s||p_0) \nonumber \\
\!\!\!\! && \!\!\!\!\!\!\!\! = \int \!\!\! q_s({\bm r}^{s}) \mathcal{F}_{C\!E}(\mathcal{D}^s|{\bm r}^{s}_k) d{\bm r}^{s} \!+\! K\!L(q_s||p_0)  \overset{\text{def}}{=}  \mathcal{L}_{1}  \label{eq:blhuc_ce}
\end{eqnarray}
where $p_{n^s}^{({\bm Y})} = \prod_{u,t} p_{n^{s}_u}^{(y^u_t)} = \prod_{u,t} p(y^u_t|\mathbb{M}_{num}^{(s,u)})$, $KL(q_s||p_0)$ denotes the KL divergence between the $q_s({\bm r}^{s})$ and prior distribution $p_0({\bm r}^{s})$, and $\mathcal{L}_{1}$ is the variational bound.
For LF-MMI criterion we use marginalization $\mathcal{F}$ to derive variational bound, which is given by
\begin{eqnarray}
\mathcal{F} & \!\!\! = \!\!\! & - \log \int \exp \{ \mathcal{F}_{LF-MMI}(\mathcal{D}^s|{\bm r}^{s})\}p_0({\bm r}^{s}) d{\bm r}^{s} \nonumber \\
& \!\!\! \leq \!\!\! & - \!\!\! \int \!\!\! q_s({\bm r}^{s}) \mathcal{F}_{L\!F\!-\!M\!M\!I}(\mathcal{D}^s|{\bm r}^{s}) d{\bm r}^{s} \!\!\!+\!\! K\!L(q_s||p_0) \overset{\text{def}}{=} \mathcal{L}_{2}.
\label{eq:blhuc_mmi}
\end{eqnarray}
For simplicity, both $q_s$ and $p_0$ are assumed to be normal distributions with $q_s({\bm r}^{s}) = \mathcal{N}({\bm \mu}_{s}, {\bm \sigma}_{s}^2)$ and $p_0({\bm r}^{s}) = \mathcal{N}({\bm \mu}_{0}, {\bm \sigma}_{0}^2)$. Hence, the $KL(q_s||p_0)$ can be computed in closed form as in \cite{xie2021bayesian}.

Considering equations (\ref{eq:mmice_lhuc}), (\ref{eq:blhuc_ce}), (\ref{eq:blhuc_mmi}), we estimate $q_s({\bm r}^{s})$ by minimizing interpolated variational bound using Monte
Carlo sampling:
\begin{eqnarray}
\mathcal{L}_{\!B\!L\!H\!U\!C} \! = \! \gamma_1\! \mathcal{L}_{2} \!+\! \gamma_2\! \mathcal{L}_{1} \!\approx\!\! \frac{1}{K} \!\! \sum\nolimits_{k=1}^K \!\! \mathcal{L}_{\!L\!H\!U\!C}(\mathcal{D}^s|{\bm r}^{s}_k) \!+\! \gamma_3\! K\!L(\!q_s\!|\!|p_0\!) \!\!\!\!\!
\label{eq:blhuc}
\end{eqnarray}
where ${\bm r}^{s}_k = {\bm \mu}_s + {\bm \sigma}_s \otimes {\bm \epsilon}_k$ and ${\bm \epsilon}_k$ is the $k$th sample drawn from standard normal distribution.
Using equation (\ref{eq:grad_lhuc}) and (\ref{eq:blhuc}) the gradients over ${\bm \mu}_{s}$ and ${\bm \sigma}_{s}$ can be derived easily. In this paper, we use $\mathcal{N}({\bm 0}, {\bm 1})$ as the prior and drawn only one sample for each update.

\subsection{Other regularization techniques}
\label{sec:reg}

To regularize parameter uncertainty, MAP \cite{huang2015maximum} and KL-regularization \cite{yu2013kl} can be applied to LHUC adaptation alternatively.
We use $\mathcal{N}({\bm 0}, {\bm 1})$ as the prior distribution for {\bf MAP-LHUC} and implemented it as L2 regularization to ${\bm r}^{s}$.
For {\bf KL-LHUC}, KL divergence between the CE output module of adapted and SI models is minimized.

To reduce the supervision uncertainty in test-time adaptation, we estimate the {\bf confidence score for each utterance} in adaptation data and select the subset with highest scores for SD parameter estimation. For this purpose, the lattice-based recognition hypotheses are exploited to compute the frame-based posteriors \cite{evermann2000posterior,mangu2000finding} of the best state sequence after initial decoding pass. The average over these posteriors serves as the confidence score for the utterance.


\section{Experiments}
\label{sec:exps}


\subsection{Experimental setup}

For the experiments, 286-hour speech collected from 4804 speakers in the \emph{Switchboard-1} corpus was used as training data. The NIST \emph{Hub5'00} and \emph{RT03} data sets containing 3.8 and 6.2 hours of speech from 80 and 144 speakers respectively were used for evaluation.
E2E LF-MMI CNN-TDNN and CNN-TDNN-BLSTM models were built in both phoneme-based and BPE-based settings. In the phoneme-based setting, 4324 full bi-phone states were used as the network output. In the fully E2E BPE-based setting, 3990 states were produced by 1995 BPE tokens.

All the models were built using a modified version of Kaldi tookit \cite{povey2011kaldi} and recipes. The CNN-TDNNs consisted of 6 ReLU activated layers of 2560-dimensional 2D-convolution with 64--256 filters, followed by 9 1536-dimensional context-splicing layers with ReLU activation, batch normalization and dropout operation. In the CNN-TDNN-BLSTMs, 10 1280-dimensional context-splicing layers were built following the 6 convolutional layers. Following each of the 6th, 8th, and 10th context-splicing layers, a BLSTM layers with 1280 cells were built. For all the models, 40-dimensional MFCC features and 100-dimensional i-vector were used as input. In the training stage, SpecAugment \cite{park2019specaugment} was applied and the training data amount was augmented to three times by using speed perturbation. 
A 4-gram LM with a 30K-word vocabulary was built with the transcripts of Switchboard and Fisher training data for decoding. Furthermore, a Transformer (TFM) LM consisting of 6 self-attention layers with 8 64-dimensional heads, and a 2-layer 2048-dimensional Bidirectional LSTM (BiLSTM) LM were built for rescoring with or without using cross-utterance in 100-word contexts \cite{sun2021transformer}.

If not specified, LHUC-based techniques were applied to the 1st and 7--12th layers in unsupervised test-time adaptation. Seven adaptation epochs were exploited with learning rate 0.1 using all the test data as adaptation data. We set $\gamma_1=1$ and $\gamma_2=0.1$ when using MMI+CE criterion, and set $\gamma_1=0$ and $\gamma_2=0.1$ when using CE only. The rate of confidence score based data selection was set to 80\% empirically. For SAT, standard LHUC adaptation was applied to the first layer and the SD and SI parameters were jointly updated.

\subsection{Results}

\renewcommand\tabcolsep{1.9pt}
\begin{table}[!htbp]
  \begin{center}
  {
    \caption{Performance of applying LHUC-based adaptation to E2E LF-MMI CNN-TDNN systems using 4-gram LM. ``SWBD'', ``CHE'', ``FSH'', and ``OA'' denote Switchboard, CallHome, Fisher, and Overall respectively. The proposed E2E adaptation framework was used if not specified, except that ``BLHUC-align'' used state alignments for adaptation. ``+conf'' denotes system using confidence score based data selection for adaptation. $\dag$ denotes the OA results with statistically significant (MAPSSWE, $\alpha=0.05$) improvement over the corresponding baseline systems.}
    \vspace{-1.0em}
    \label{tab:cnn0}
    \scalebox{0.62}[0.65]
    {
    \begin{tabular}{c|c l||c l c|c c c||c l c|c c c} \toprule[0.7pt]
         \multirow{3}{*}{id} & \multirow{2}{*}{Adapt} & \multirow{2}{*}{Adapt} & \multicolumn{6}{c||}{WER (\%) in {\bf Phoneme}-based setting} & \multicolumn{6}{c}{WER (\%) in {\bf BPE}-based setting} \\ \cline{4-15}
          & \multirow{2}{*}{Obj} & \multirow{2}{*}{Method} & \multicolumn{3}{c|}{\emph{Hub5'00}} & \multicolumn{3}{c||}{\emph{RT03}} & \multicolumn{3}{c|}{\emph{Hub5'00}} & \multicolumn{3}{c}{\emph{RT03}} \\ \cline{4-15}
           & & & \emph{SWBD} & \emph{CHE} & \multicolumn{1}{c|}{OA} & \emph{SWBD} & \emph{FSH} & OA & \emph{SWBD} & \emph{CHE} & \multicolumn{1}{c|}{OA} & \emph{SWBD} & \emph{FSH} & OA \\ \toprule[0.7pt]
         1& - & - & 8.8 & 17.0 & 12.9 & 19.2 & 11.5 & 15.5 & 9.1 & 17.3 & 13.2 & 18.8 & 12.0 & 15.5 \\ \toprule[0.7pt]
         2& \multirow{4}{*}{MMI} & LHUC-oracle & 6.5 & 10.6 & 8.6$\dag$ & 12.1 & 7.5 & 9.9$\dag$ & 8.2 & 14.1 & 11.2$\dag$ & 14.9 & 9.7 & 12.4$\dag$ \\ \cline{3-15}
         3& \multirow{4}{*}{+CE} & LHUC & 8.7 & 16.9 & 12.8 & 19.1 & 11.5 & 15.4 & 8.8 & 16.5 & 12.7$\dag$ & 18.2 & 11.6 & 15.0$\dag$ \\
         4& & LHUC+conf & \textbf{8.4} & 16.5 & 12.5$\dag$ & 18.8 & 11.3 & 15.2$\dag$ & 8.8 & 16.4 & 12.6$\dag$ & 18.1 & 11.4 & 14.9$\dag$ \\
         5& & BLHUC & 8.5 & 16.0 & 12.3$\dag$ & 17.9 & 10.9 & 14.5$\dag$ & 8.9 & 16.2 & 12.6$\dag$ & 17.6 & 11.1 & 14.5$\dag$ \\
         6& & BLHUC+conf & \textbf{8.4} & \textbf{15.9} & \textbf{12.2}$\dag$ & \textbf{17.8} & \textbf{10.7} & \textbf{14.4}$\dag$ & \textbf{8.7} & \textbf{16.0} & \textbf{12.4}$\dag$ & \textbf{17.5} & \textbf{10.9} & \textbf{14.3}$\dag$ \\ \toprule[0.7pt]
         7& \multirow{10}{*}{CE} & LHUC-oracle & 7.4 & 13.1 & 10.3$\dag$ & 14.3 & 9.0 & 11.7$\dag$ & 8.2 & 14.9 & 11.6$\dag$ & 15.0 & 10.1 & 12.6$\dag$ \\ \cline{3-15}
         8& & LHUC & 8.6 & 16.2 & 12.4$\dag$ & 18.4 & 11.2 & 14.9$\dag$ & \textbf{8.7} & 16.1 & 12.4$\dag$ & 18.0 & 11.4 & 14.8$\dag$ \\
         9& & LHUC+conf & 8.5 & 16.1 & 12.3$\dag$ & 18.4 & 11.1 & 14.9$\dag$ & 8.8 & 16.0 & 12.4$\dag$ & 18.1 & 11.3 & 14.8$\dag$ \\
         10& & KL-LHUC & 8.5 & 16.2 & 12.4$\dag$ & 18.4 & 11.2 & 14.9$\dag$ & \textbf{8.7} & 16.2 & 12.5$\dag$ & 18.2 & 11.5 & 15.0$\dag$ \\
         11& & MAP-LHUC & 8.5 & 16.1 & 12.3$\dag$ & 18.4 & 11.3 & 15.0$\dag$ & 8.8 & 16.1 & 12.5$\dag$ & 18.0 & 11.4 & 14.8$\dag$ \\
         12& & BLHUC & \textbf{8.3} & \textbf{15.5} & \textbf{11.9}$\dag$ & \textbf{17.4} & 10.7 & \textbf{14.2}$\dag$ & \textbf{8.7} & \textbf{15.7} & \textbf{12.2}$\dag$ & \textbf{17.5} & 11.0 & \textbf{14.4}$\dag$ \\
         13& & BLHUC+conf & \textbf{8.3} & 15.8 & 12.1$\dag$ & 17.8 & \textbf{10.6} & 14.3$\dag$ & 8.7 & 15.8 & 12.3$\dag$ & 17.8 & \textbf{10.9} & 14.5$\dag$ \\
         14& & BLHUC-align & 8.7 & 16.2 & 12.5$\dag$ & 18.1 & 11.1 & 14.7$\dag$ & 8.9 & 16.4 & 12.7$\dag$ & 17.7 & 11.4 & 14.7$\dag$ \\ \cline{3-15}
         15& & SAT+LHUC & 8.3 & 15.3 & 11.8$\dag$ & 17.8 & 10.8 & 14.4$\dag$ & 8.5 & \textbf{15.4} & 12.0$\dag$ & 17.5 & 11.0 & 14.4$\dag$ \\
         16& & SAT+BLHUC & \textbf{8.1} & \textbf{15.0} & \textbf{11.6}$\dag$ & \textbf{16.8} & \textbf{10.4} & \textbf{13.7}$\dag$ & \textbf{8.3} & 15.5 & \textbf{11.9}$\dag$ & \textbf{17.3} & \textbf{10.7} & \textbf{14.1}$\dag$ \\
          \toprule[0.7pt]
    \end{tabular}
    }
  }
  \end{center}
\end{table}

Performance of applying different LHUC-based adaptation techniques with MMI+CE or CE criterions to the E2E LF-MMI TDNN system built with either phoneme-based or BPE-based setting and using 4-gram LM is shown in Table \ref{tab:cnn0}. The proposed E2E adaptation framework was used if not specified.
Four main trends can be found. First, when using ground true word sequences for supervision, the oracle adaptation methods using MMI+CE criterion produced better performance than using CE criterion only (id 2 vs 7). In contrast, the unsupervised adaptation methods with CE criterion consistently outperformed those with MMI+CE criterion (e.g., id 8 vs 3, and id 12 vs 5). This may be due to the supervision errors (with up to 21\% of mono-phone error rate and 35\% of BPE error rate) in unsupervised adaptation, and the discriminative LF-MMI criterion requires more accurate supervision labels.
Second, consistent improvement was obtained by applying confidence score based data selection to the adaptation methods with MMI+CE (ids 4 and 6), which may be caused by the slight reduction (about 2\% in absolute) of supervision errors with data selection. However, these results were still worse than the results of adaptation using CE only, and no improvement was obtained by data selection when using CE only (ids 9 and 13).
\renewcommand\tabcolsep{1.4pt}
\begin{table}[!h]
  \begin{center}
  {
    \caption{Performance of BLHUC with CE in E2E adaptation framework for E2E LF-MMI CNN-TDNN and CNN-TDNN-BLSTM. }
    \vspace{-1.0em}
    \label{tab:cnn}
    \scalebox{0.56}[0.6]
    {
    \begin{tabular}{c|c c l||c l c|c c c||c l c|c c c} \toprule[0.7pt]
     \multirow{3}{*}{id} & \multirow{1}{*}{E2E} & \multirow{3}{*}{LM} & \multirow{2}{*}{Adapt} & \multicolumn{6}{c||}{WER (\%) in {\bf Phoneme}-based setting} & \multicolumn{6}{c}{WER (\%) in {\bf BPE}-based setting} \\ \cline{5-16}
          & \multirow{1}{*}{LF-MMI} & & \multirow{2}{*}{Method} & \multicolumn{3}{c|}{\emph{Hub5'00}} & \multicolumn{3}{c||}{\emph{RT03}} & \multicolumn{3}{c|}{\emph{Hub5'00}} & \multicolumn{3}{c}{\emph{RT03}} \\ \cline{5-16}
          & \multirow{1}{*}{Network} & & & \emph{SWBD} & \emph{CHE} & \multicolumn{1}{c|}{OA} & \emph{SWBD} & \emph{FSH} & OA & \emph{SWBD} & \emph{CHE} & \multicolumn{1}{c|}{OA} & \emph{SWBD} & \emph{FSH} & OA \\ \toprule[0.7pt]
         1& \multirow{8}{*}{CNN-} & \multirow{3}{*}{4gram} & - & 8.8 & 17.0 & 12.9 & 19.2 & 11.5 & 15.5 & 9.1 & 17.3 & 13.2 & 18.8 & 12.0 & 15.5 \\
         2& \multirow{8}{*}{TDNN} & & BLHUC & 8.3 & 15.5 & 11.9$\dag$ & 17.4 & 10.7 & 14.2$\dag$ & 8.7 & 15.7 & 12.2$\dag$ & 17.5 & 11.0 & 14.4$\dag$ \\
         3& & & SAT+BLHUC & {\bf 8.1} & {\bf 15.0} & {\bf 11.6}$\dag$ & {\bf 16.8} & {\bf 10.4} & {\bf 13.7}$\dag$ & {\bf 8.3} & {\bf 15.5} & {\bf 11.9}$\dag$ & {\bf 17.3} & {\bf 10.7} & {\bf 14.1}$\dag$ \\ \cline{3-16}
         4& & \multirow{3}{*}{+TFM} & - & 7.4 & 14.9 & 11.2 & 16.6 & 9.6 & 13.2 & 7.7 & 15.3 & 11.5 & 16.4 & 10.1 & 13.4 \\
         5& & & BLHUC & 6.9 & 13.2 & 10.1$\dag$ & 15.0 & 8.7 & 12.0$\dag$ & {\bf 7.1} & 14.0 & 10.6$\dag$ & 15.1 & 9.1 & 12.2$\dag$ \\
         6& & & SAT+BLHUC & \textbf{6.8} & \textbf{13.0} & \textbf{9.9}$\dag$ & \textbf{14.7} & \textbf{8.4} & \textbf{11.7}$\dag$ & {\bf 7.1} & {\bf 13.7} & {\bf 10.4}$\dag$ & {\bf 15.0} & {\bf 8.9} & {\bf 12.1}$\dag$ \\ \cline{3-16}
         7& & \multirow{2}{*}{++BiLSTM} & - & 6.6 & 13.3 & 10.0 & 14.9 & 8.5 & 11.8 & 6.8 & 14.0 & 10.4 & 14.5 & 9.0 & 11.8 \\
         8& & \multirow{2}{*}{+cross-utt} & BLHUC & 6.2 & 12.1 & 9.2$\dag$ & 13.2 & 7.7 & 10.5$\dag$ & \textbf{6.4} & 12.8 & 9.6$\dag$ & 13.6 & 7.9 & 10.9$\dag$ \\
         9& & & SAT+BLHUC & \textbf{6.1} & \textbf{11.6} & \textbf{8.9}$\dag$ & \textbf{13.0} & \textbf{7.4} & \textbf{10.3}$\dag$ & \textbf{6.4} & \textbf{12.4} & \textbf{9.4}$\dag$ & \textbf{13.4} & \textbf{7.8} & \textbf{10.7}$\dag$ \\
          \toprule[0.7pt]
         10& \multirow{7}{*}{CNN-} & \multirow{3}{*}{4gram} & - & 8.8 & 16.4 & 12.6 & 17.7 & 11.0 & 14.5 & 8.7 & 15.9 & 12.3 & 17.6 & 11.2 & 14.5 \\
         11& \multirow{7}{*}{TDNN-} & & BLHUC & 8.5 & 15.3 & 11.9$\dag$ & 16.5 & 10.2 & 13.5$\dag$ & 8.4 & 14.8 & 11.6$\dag$ & 16.3 & 10.1 & 13.3$\dag$ \\
         12& \multirow{7}{*}{BLSTM} & & SAT+BLHUC & \textbf{8.3} & \textbf{14.6} & \textbf{11.5}$\dag$ & \textbf{16.0} & \textbf{9.9} & \textbf{13.1}$\dag$ & \textbf{8.0} & \textbf{14.2} & \textbf{11.1}$\dag$ & \textbf{15.4} & \textbf{9.5} & \textbf{12.6}$\dag$ \\ \cline{3-16}
         13& & \multirow{3}{*}{+TFM} & - & 7.0 & 14.2 & 10.6 & 15.5 & 9.0 & 12.4 & 7.3 & 14.2 & 10.8 & 15.5 & 9.2 & 12.5 \\
         14& & & BLHUC & \textbf{6.9} & 13.1 & 10.0$\dag$ & 14.3 & 8.4 & 11.5$\dag$ & 7.1 & 13.2 & 10.2$\dag$ & 14.2 & 8.3 & 11.4$\dag$ \\
         15& & & SAT+BLHUC & \textbf{6.9} & \textbf{12.9} & \textbf{9.9}$\dag$ & \textbf{13.9} & \textbf{8.3} & \textbf{11.2}$\dag$ & \textbf{6.8} & \textbf{12.7} & \textbf{9.8}$\dag$ & \textbf{13.4} & \textbf{8.1} & \textbf{10.8}$\dag$ \\ \cline{3-16}
         16& & \multirow{2}{*}{++BiLSTM} & - & 6.5 & 12.9 & 9.7 & 13.9 & 8.1 & 11.1 & 6.7 & 13.3 & 10.0 & 14.1 & 8.4 & 11.4 \\
         17& & \multirow{2}{*}{+cross-utt} & BLHUC & 6.4 & 12.1 & 9.3$\dag$ & 12.8 & 7.5 & 10.2$\dag$ & 6.5 & 12.3 & 9.4$\dag$ & 12.7 & 7.7 & 10.3$\dag$ \\
         18& & & SAT+BLHUC & \textbf{6.2} & \textbf{11.7} & \textbf{9.0}$\dag$ & \textbf{12.5} & \textbf{7.4} & \textbf{10.0}$\dag$  & \textbf{6.2} & \textbf{11.7} & \textbf{9.0}$\dag$ & \textbf{12.0} & \textbf{7.2} & \textbf{9.7}$\dag$ \\
          \toprule[0.7pt]
    \end{tabular}
    }
  }
  \end{center}
\end{table}
Third, systems applying BLHUC adaptation in E2E framework with CE criterion (id 12) consistently outperformed the standard LHUC adapted systems (id 8), baseline systems (id 1), as well as the systems adapted by BLHUC using state alignments (id 14). No improvement was produced by using MAP-LHUC or KL-LHUC adaptation compared to the standard LHUC adaptation.
Finally, joint use of SAT and LHUC/BLHUC adaptation with CE criterion (ids 15 and 16) achieved significant further improvement over using test-time LHUC/BLHUC adaptation only, and outperformed the baseline systems (id 1) by relative WER reductions up to \textbf{10.5\%} and \textbf{11.4\%} on Hub5'00 and RT03 evaluation sets, respectively.

\vspace{0.8em}
\begin{figure}[!htbp]
  \begin{center}
    \includegraphics[width=8cm,height=2cm]{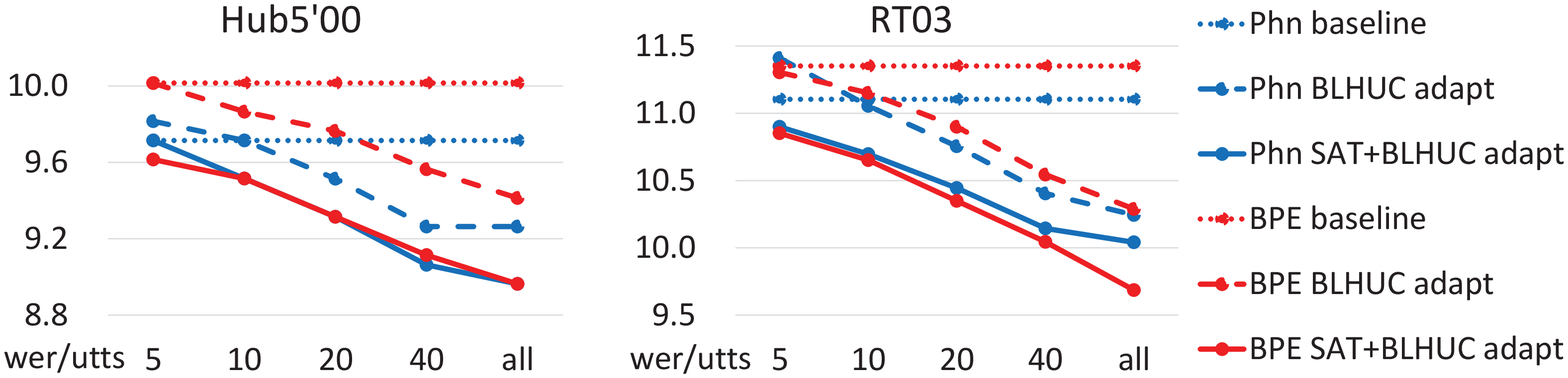}
    \vspace{-1.0em}
    \caption{BLHUC adaptation using varying number of utterances for each speaker on the best BLSTM systems (ids 16--18 in Table \ref{tab:cnn}). }
    \label{fig:amount}
  \end{center}
\end{figure}

Table \ref{tab:cnn} shows the performance of applying BLHUC with CE criterion in E2E adaptation framework to E2E LF-MMI CNN-TDNN and CNN-TDNN-BLSTM models.
Consistent to the Table \ref{tab:cnn0}, the BLHUC or SAT+BLHUC adapted CNN-TDNN and CNN-TDNN-BLSTM systems consistently and significantly outperformed the corresponding baseline systems with or without using neural network LM rescoring,
by relative WER reductions up to \textbf{11.1\%} and \textbf{12.8\%} on Hub5'00 and RT03 in phoneme-based setting (id 9 vs 7), and \textbf{10.5\%} and \textbf{14.7\%} in BPE-based setting (id 18 vs 16), respectively.
Fig \ref{fig:amount} shows the performance of BLHUC adaptation using varying adaptation data amounts on the CNN-TDNN-BLSTM systems with TFM+BiLSTM rescoring (ids 16--18).

\renewcommand\tabcolsep{2pt}
\begin{table}[!htbp]
  \begin{center}
  {
    \caption{Performance contrasts of the best adapted E2E LF-MMI systems against other SOTA systems on the 300-hour SWBD setup. }
    \vspace{-1.0em}
    \label{tab:compare}
    \scalebox{0.62}[0.6]
    {
    \begin{tabular}{c|l|c|c c c|c c c} \toprule[0.7pt]
          \multirow{2}{*}{id} & \multirow{2}{*}{System} & \multirow{2}{*}{\#Param} & \multicolumn{3}{c|}{WER (\%) of \emph{Hub5'00}} & \multicolumn{3}{c}{WER (\%) of \emph{RT03}} \\ \cline{4-9}
          & & & \emph{SWBD} & \emph{CHE} & \multicolumn{1}{c|}{OA} & \emph{SWBD} & \emph{FSH} & OA \\ \toprule[0.7pt]
          1& RWTH-2019 Hybrid BLSTM + Adapt \cite{kitza2019cumulative} & - & 6.7 & 13.5 & 10.2 & - & - & - \\
          2& CUHK-2021 Hybrid LF-MMI + Adapt \cite{xie2021bayesian} & 15M & 6.7 & 12.7 & 9.7 & 13.4 & 7.9 & 10.7 \\ \hline
          3& Google-2019 LAS \cite{park2019specaugment} & - & 6.8 & 14.1 & (10.5) & - & - & - \\
          4& IBM-2020 BPE AED \cite{tuske2020single} & 280M & 6.4 & 12.5 & (9.5) & 14.8 & 8.4 & (11.7) \\
          5& Salesforce-2020 Phn\&BPE TFM \cite{wang2020investigation} & - & 6.3 & 13.3 & (9.8) & - & - & 11.4 \\
          6& IBM-2021 BPE CFM-AED + TFMXL-LM \cite{tuske2021limit} & 68M & \textbf{5.5} & \textbf{11.2} & \textbf{(8.4)} & 12.6 & \textbf{7.0} & (9.9) \\
          7& CUHK-CAS-2022 BPE CFM + Adapt \cite{deng2022confidence} & 46M & 6.3 & 13.0 & 9.6 & 13.8 & 8.7 & 11.3 \\ \toprule[0.7pt]
          8& \textbf{Phn E2E LF-MMI CNN-TDNN (ours)} & 14M & 6.6 & 13.3 & 10.0 & 14.9 & 8.5 & 11.8 \\
          9&  \textbf{+ Adapt (ours, id 9 in Table \ref{tab:cnn})} & +12k & 6.1 & 11.6 & 8.9 & 13.0 & 7.4 & 10.3 \\
          10& \textbf{Phn E2E LF-MMI CNN-TDNN-BLSTM (ours)} & 48M & 6.5 & 12.9 & 9.7 & 13.9 & 8.1 & 11.1 \\
          11&  \textbf{+ Adapt (ours, id 18 in Table \ref{tab:cnn})} & +10k & 6.2 & 11.7 & 9.0 & 12.5 & 7.4 & 10.0  \\ \hline
          12& \textbf{BPE E2E LF-MMI CNN-TDNN (ours)} & 14M & 6.8 & 14.0 & 10.4 & 14.5 & 9.0 & 11.8 \\
          13&  \textbf{+ Adapt (ours, id 9 in Table \ref{tab:cnn})} & +12k & 6.4 & 12.4 & 9.4 & 13.4 & 7.8 & 10.7  \\
          14& \textbf{BPE E2E LF-MMI CNN-TDNN-BLSTM (ours)} & 48M & 6.7 & 13.3 & 10.0 & 14.1 & 8.4 & 11.4 \\
          15&  \textbf{+ Adapt (ours, id 18 in Table \ref{tab:cnn})} & +10k & 6.2 & 11.7 & 9.0 & \textbf{12.0} & 7.2 & \textbf{9.7} \\ \toprule[0.7pt]
    \end{tabular}
    }
  }
  \end{center}
\end{table}

Table \ref{tab:compare} compares the results of the best adapted E2E LF-MMI systems to the state-of-the-art (SOTA) results on the same task obtained by the most recent hybrid and end-to-end systems without system combination reported in the literature. In particular, the best adapted BPE-based E2E LF-MMI CNN-TDNN-BLSTM system outperformed the SOTA results (id 15 vs 6) on the RT03 task.

\section{Conclusion}
\label{sec:concl}

This paper investigates the LHUC based adaptation techniques applied to E2E LF-MMI models in phoneme-based and BPE-based settings. An unsupervised E2E model-based adaptation framework is proposed for SD parameter estimation using LF-MMI and CE criterions. Moreover, BLHUC is employed to mitigate the risk of overfitting during adaptation on E2E LF-MMI CNN-TDNN and CNN-TDNN-BLSTM models.
Experiments on the 300-hour Switchboard task suggest that applying BLHUC in the proposed E2E adaptation framework to E2E LF-MMI systems consistently outperformed the baseline systems by relative WER reductions up to 11.1\% and 14.7\% on the Hub5'00 and RT03, and achieved the best performance in WERs of 9.0\% and 9.7\%, respectively. These are comparable to the SOTA adapted LF-MMI hybrid systems and the SOTA adapted Conformer-based E2E systems. Future work will focus on rapid on-the-fly adaptation approaches for E2E models.


\ninept

\bibliographystyle{IEEEbib}
\bibliography{adapt_lda}

\end{document}